\documentclass[aps,twocolumn]{revtex4}
\makeatletter

\newcommand{\Rmnum}[1]{\expandafter\@slowromancap\romannumeral #1@}
\makeatother
\usepackage[titletoc]{appendix}
\usepackage{graphicx}
\usepackage{epstopdf}
\usepackage{dcolumn}
\usepackage{bm}
\usepackage{CJK}
\usepackage{amsfonts}
\usepackage{psfrag}
\usepackage{wrapfig}
\usepackage{subfigure}
\usepackage{makeidx}
\usepackage{color, soul}
\usepackage[colorlinks,linkcolor=red,anchorcolor=red,citecolor=blue,urlcolor=blue]{hyperref}
\usepackage{multirow}
\usepackage{epsf}
\usepackage{appendix}
\usepackage{hyperref}
\usepackage{amsmath}
\usepackage{cases}\tolerance=1
\emergencystretch=\maxdimen
\begin{document}
\title{Breather induced quantized superfluid vortex filaments and their characterization}

\author{Hao Li$^{1,2}$}
\author{Chong Liu$^{1,2}$}\email{chongliu@nwu.edu.cn}
\author{Wei Zhao$^{1,4}$}
\author{Zhan-Ying Yang$^{1,2}$}\email{zyyang@nwu.edu.cn}
\author{Wen-Li Yang$^{1,2,3}$}
\address{$^1$School of Physics, Northwest University, Xi'an 710069, China}
\address{$^2$Shaanxi Key Laboratory for Theoretical Physics Frontiers, Xi'an 710069, China}
\address{$^3$Institute of Modern Physics, Northwest University, Xi'an 710069, China}
\address{$^4$Institute of Photonics and Photon-technology, Northwest University, Xi'an 710069, China}

\date{\today}
\begin{abstract}
We study and characterize the breather-induced quantized superfluid vortex filaments which correspond to the Kuznetsov-Ma breather and super-regular breather excitations developing from localised perturbations. Such vortex filaments, emerging from an otherwise perturbed helical vortex, exhibit intriguing loop structures corresponding to the large amplitude of breathers due to the dual action of bending and twisting of the vortex.
The loop induced by Kuznetsov-Ma breather emerges periodically as time increases, while the loop structure triggered by super-regular breather---\textit{the loop pair}---exhibits striking symmetry breaking due to the broken reflection symmetry of the group velocities of super-regular breather.
In particular, we identify explicitly the generation conditions of these loop excitations by introducing a physical quantity---the integral of the relative quadratic curvature---which corresponds to the effective energy of breathers. Although the nature of nonlinearity, it is demonstrated that this physical quantity shows a \textit{linear} correlation with the loop size.
These results will deepen our understanding of breather-induced vortex filaments and be helpful for controllable ring-like excitations on the vortices.
\end{abstract}

\pacs{}

\maketitle

\section{INTRODUCTION}

Quantum fluid \cite{superfluid1,quantumfluid} has recently been the subject of
extensive investigations that contains vortex generation, interaction, and reconnection of vortex lines influenced by vortices
\cite{vortex reconnection,vortices1}. The motion of quantum fluid is most succinctly illustrated by vortex filament due that it consists of vorticity of infinite strength concentrated along the filament and gives an intuitive geometric interpretation of the evolution of the vorticity field. In the case of ideal inviscid fluid, the motion of the fluid elements is constrained by Biot-Savart law which provides valuable information of vortex tangles \cite{BS,vortex tangles1}. In particular, a variety of excitations are generated and evolve along the vortex filament due to the self-induced velocity \cite{LIA2,AB,multibreather,excitations1,excitations2}. Such excitations are physically important since they turn out to be the main degrees of freedom remaining in a superfluid in the ultra low temperature regime.
Therefore, investigation of vortex structures of different fundamental excitations is both relevant and necessary.

The first prototype of such excitations is the so-called `Kelvin wave' \cite{KW1}. The latter, which is originated from the small deformations of vortex lines, plays an important role in the decay of turbulence energy \cite{KW2}. However, one should note that Kelvin waves are low amplitude linear excitations of a straight vortex. In contrast, there are also some larger amplitude excitations propagating along the filament. Such large-amplitude excitations are induced by nonlinearity, i.e., `nonlinear excitations'.

There exists an interesting link between nonlinear excitations and vortex dynamics that has attracted considerable attention recently.
It is presently known as `Hasimoto transformation' \cite{LIA2} that allows us map the motion of vortex filament onto a scalar cubic nonlinear Schr\"{o}dinger equation (NLSE) of the self-focusing type and the resulting loop structure of bright soliton on a vortex filament is demonstrated  \cite{LIA2}.
In addition to the classical solitons, the NLSE possesses rich `breathing' excitations on a plane wave background which are known as `breathers' \cite{Book97,O1a,SR1}.
Such breathers are strongly associated with the modulation instability (MI) \cite{Book97,O1a,SR1,OE,MI1}, where its nonlinear stage has been regarded as the prototype of rogue wave events \cite{Extreme09}. Surprisingly, although breather has been one of the center subjects in nonlinear physics
and its observation has been realized widely in many nonlinear systems \cite{O1a,W1b,O1b,Onorato,Wabnitz}, the link between one special type of breather---Akhmediev breather \cite{ABbreather} (as well as its multiple counterpart) and vortex filaments in a quantized superfluid has been revealed only recently \cite{AB,multibreather}. In fact, the resulting new loop structure of Akhmediev breather, which differs from that of bright solitons \cite{LIA2}, provides significant contributions to our understanding of quantum fluid and superfluid turbulence.
This is therefore an interdisciplinary
research---in the case of quantized superfluid vortex filaments, MI, and breathers---that
needs more explorations.


However, the Akhmediev breather is merely the exact description for the MI emerging from a special purely periodic perturbations \cite{ABbreather}.
There is another type of breathers describing the MI developing from localised perturbations that has not been studied in a quantized superfluid. This includes the Kuznetsov-Ma breather \cite{KMB} admitting localised single-peak perturbation and super-regular breather \cite{SR1} supporting localised multi-peak perturbation. Indeed, it is recently demonstrated that Kuznetsov-Ma breather describes not only the MI in the small amplitude regime but also the interference between bright soliton and plane wave in the large amplitude regime \cite{Mechanism of KMB}; while the super-regular breather admits the MI growth rate that coincides with the absolute difference of group velocities of the breather \cite{SR3,SR4}. Given that these breathers are qualitatively different, two questions of fundamental importance now arise: How about the loop excitations triggered by these breathers? Is there a physical quantity to identify explicitly all these breather-induced loop excitations?




In this paper, we study the quantized superfluid vortex filaments induced by Kuznetsov-Ma breather and super-regular breather admitting localised perturbations. Such vortex filaments exhibit striking loop structures due to the dual action of bending and twisting of the vortex. Remarkably, an intriguing loop structure triggered by super-regular breather---the loop pair---exhibits spontaneous symmetry breaking, due to the broken reflection symmetry of the group velocities of super-regular breather.
In particular, we identify explicitly these loop excitations by introducing the integral of the relative quadratic curvature, which corresponds to the effective energy of breathers. Although the nature of nonlinearity, it is demonstrated that this physical quantity shows a linear correlation with the loop size.

\section{Hasimoto transformation and inverse map}
For the incompressible and inviscid fluid, the Biot-Savart equation is reduced to a simpler local induction approximation (LIA) equation \cite{LIA1,liq3,liq4} by taking leading order
\begin{eqnarray}
\label{equ:LIA}
\mathbf{v}=\left(\Gamma/4\pi\right)\ln\left(R/a_0\right)\kappa \mathbf{t}\times\mathbf{n}=\beta \kappa \mathbf{t}\times\mathbf{n}.
\end{eqnarray}
Here $\Gamma$ is a circulation, $R$ is local radius of curvature and $a_0$ is the effective vortex core radius. $\mathbf{v}=\frac{d\mathbf{r}}{dt}$ is the flow velocity vector of the vortex filament, $\mathbf{t}$ and $\mathbf{n}$ are unit vectors corresponding to the tangent and principal normal directions, respectively. $\kappa$, as a real function of arc length variable $s$ and time $t$, represents the curvature distribution of the vortex filament. This equation makes us obtain more properties of the states related to the motion of vortex especially in the case of Hasimoto transformation. Assuming that $\beta$ is constant and making use of the Seret-Frenet equations given in \cite{SFe},
\begin{eqnarray}
\label{equ:2}
\mathbf{r}'=\mathbf{t},~~\mathbf{t}'=\kappa\mathbf{n},~~\mathbf{n}'=\tau\mathbf{b}-\kappa\mathbf{t},~~\mathbf{b}'=-\tau\mathbf{n},
\end{eqnarray}
where prime denotes a differential of arc length, $\mathbf{b}$ is binormal vector and $\tau$ is the torsion of the vortex filament, Eq. (\ref{equ:LIA}) can be transformed into a 1D scalar cubic NLSE of self-focusing type \cite{LIA2}
\begin{eqnarray}
\label{equ:NLSE}
\beta^{-1}\left(i\psi_t\right)=-\psi_{ss}-\frac{1}{2}|\psi|^2\psi.
\end{eqnarray}
$\psi\left(s,t\right)$ is a complex function related with the local instantaneous geometric parameters curvature $\kappa\left(s,t\right)$ and torsion $\tau\left(s,t\right)$ in the context of vortices by the transformation
\begin{eqnarray}
\psi\left(s,t\right)=\kappa\left(\sigma,t\right)e^{i\int_0^s \tau\left(\sigma,t\right)ds}.
\end{eqnarray}

The NLSE (\ref{equ:NLSE}) possesses rich `breathing' excitations \cite{Book97}, which provides a path for studying exactly breather-induced quantized superfluid vortex filaments. One can obtain the explicit configuration of these excitations by inverse map (see Appendix \ref{Appendix A}).

\section{Kuznetsov-Ma breather induced vortex filaments and exact characterization}\label{Sec3}
We first consider the Kuznetsov-Ma breather that exhibits periodic pulsating dynamics along $t$. Its explicit expression for Eq. (\ref{equ:NLSE}) is given by
\begin{eqnarray}
\label{equ:KMB}                                                                                                                                                                                                                                                                \psi(s,t)=\left[1-2\frac{\chi^2\cos\left(\eta\beta t\right)+i\eta \sin\left(\eta\beta t\right)}{\kappa_0b~\cosh\left(\chi\xi\right)-\kappa_0^2~\cos\left(\eta\beta t\right)}\right]\psi_0,
\end{eqnarray}
where $\chi=\sqrt{b^2-\kappa_0^2}$ with $b$ being a real constant ($b>\kappa_0$), $\eta=b\chi$, and $\xi=s-2\tau_0\beta t$. Physically, $b$ describes the oscillation period and amplitude of the Kuznetsov-Ma breather. $\kappa_0$ and $\tau_0$ are real constants which denote the amplitude and wave vector of
the plane wave background $\psi_0$ respectively. The latter has the from
\begin{eqnarray}
\label{equ:PWB}                                                                                                                                                                                                                                                                  &&\psi_0=\kappa_0\exp{i(\tau_0s+\omega t)},~\omega=\beta\,\kappa_0^2/2-\beta\,\tau_0^2.
\end{eqnarray}
This plane wave corresponds to a trivial uniform helical vortex without physical interest ($\kappa_0$ and $\tau_0$ describe the curvature and torsion of a uniform helical vortex, respectively). In contrast, the Kuznetsov-Ma breather describes nontrivial structure of vortex filament that has not been studied fully. From Eq. (\ref{equ:KMB}), one can readily calculate the explicit expressions of the curvature and torsion of vortex filament induced by Kuznetsov-Ma breather, which are given respectively by
\begin{eqnarray}
\label{equ:11}
\kappa=\left[\left(\kappa_0+\frac{2\chi^2\cos(\eta\beta t)}{n_1}\right)^2+\frac{4\eta^2\sin^2(\eta\beta t)}{\left(n_1\right)^2}\right]^{1/2},
\end{eqnarray}
and
\begin{eqnarray}
\label{equ:12}
\tau=\tau_0 \left[1+\frac{4\kappa_0\eta^2\sin\left(\eta\beta t\right)\sinh\left(\chi\xi\right)}{m_1+m_2+m_3+m_4}\right],
\end{eqnarray}
with
\begin{eqnarray}
\nonumber
&&n_1=\kappa_0\cos\left(\eta\beta t\right)-b\cosh\left(\chi\xi\right),\\\nonumber
&&m_1=\kappa_0^4-7\kappa0^2b^2+8b^4,~m_2=\kappa_0^4\cos\left(2\eta\beta t\right),\\\nonumber
&&m_3=4\kappa_0 b\left(\kappa_0^2-2b^2\right)\cos\left(\eta\beta t\right)\cosh\left(\chi\xi\right),\\\nonumber
&&m_4=\kappa_0^2b^2\cosh\left(2\chi\xi\right).
\end{eqnarray}

\begin{figure}[htbp]
  \centering
  \includegraphics[width=82mm]{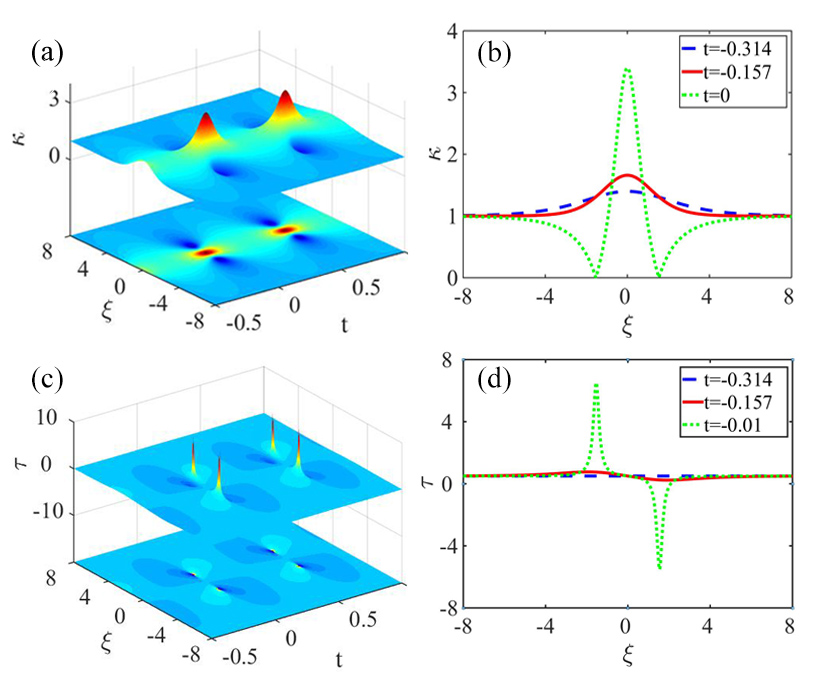}\\
  \caption{Temporal evolutions of curvature $\kappa\left(\xi,t\right)$ (a) and torsion $\tau\left(\xi,t\right)$ (c) corresponding to a Kuznetsov-Ma breather, see Eqs. (\ref{equ:11}) and (\ref{equ:12}). (b) and (d) are variations of $\kappa\left(\xi,t\right)$ and $\tau\left(\xi,t\right)$ at different times.  The parameters are: $\kappa_0=1$, $\tau_0=0.05$, $b=1.2$, and $\beta=4\pi$.}\label{fig1}
\end{figure}

The variations of curvature and torsion of Kuznetsov-Ma breather on the ($\xi$, $t$) plane (note here that $\xi=s-2\tau_0\beta t$ denotes the moving frame on the group velocity), with initial conditions $b=1.2$, $\kappa_0=1$, $\tau_0=0.05$, are shown in Fig. \ref{fig1}(a) and (c).
As expected, the curvature of Kuznetsov-Ma breather, starting from a localised non-periodic (single-peak) perturbation, evolves gradually into its maximum at $t=0$ [see the profiles in Fig. \ref{fig1}(b)].
The curvature then exhibits periodic oscillation with the period $2\pi/(\eta\beta)$ as $t$ increases [see Fig. \ref{fig1}(a)].

A notable feature is that the torsion, as a function of arc length $s$ and time $t$, exhibits singular behavior as $t\rightarrow 2\pi/(\eta\beta)$. Figure \ref{fig1}(d) clearly indicates that the phase becomes ill defined at the point $\kappa=0$ near $t=0$, which leads to the severe twisting of the vortex filament. This is not surprising since the Kuznetsov-Ma breather admits a $\pi$ phase shift at the valleys. This phase shift results in the singular behavior of the torsion. Note that the singular does not make the Hasimoto transformation ill-defined due to the $\pi$ phase shift of nonlinear waves.

\begin{figure*}[htbp]
  \centering
  \includegraphics[width=150mm]{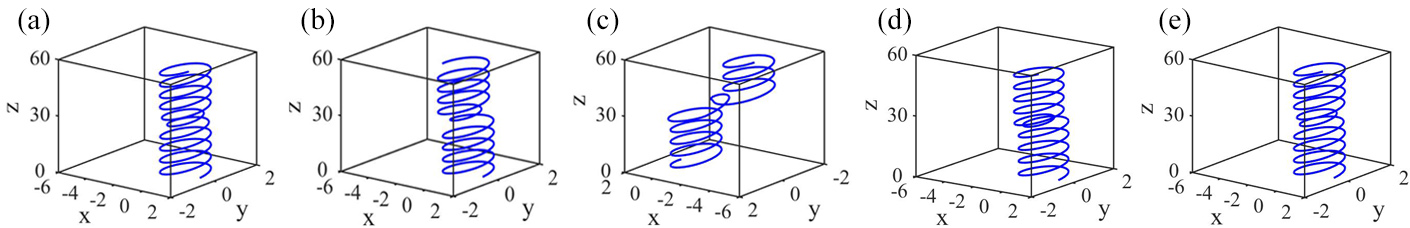}\\
  \caption{Configuration of vortex filaments of Kuznetsov-Ma breathers computed by LIA for parameters given by $\kappa_0=1$, $\tau_0=0.05$, $b=1.2$ and $\beta=4\pi$ at different time (a) $t=-0.314$, (b) $t=-0.157$, (c) $t=-0.01$, (d) $t=0.157$ and (e) $t=0.314$.}\label{fig2}
\end{figure*}

Figure \ref{fig2} shows the corresponding vortex configuration of Kuznetsov-Ma breather within one growth-decay cycle.
One can see clearly from the figure that the vortex filament, emerging from an otherwise perturbed helical vortex at $t=-0.314$ [see Fig. \ref{fig2} (a)], exhibits a striking loop structure at $t=0$ due to the dual action of bending and twisting of the vortex. This loop structure disappears gradually as $t$ increases.
At $t=0.314$, the vortex filament recovers the initial state. This process will emerge periodically as $t$ increases due to the feature of the Kuznetsov-Ma breather. One should note that when $\kappa_0\rightarrow0$, the loop of Kuznetsov-Ma breather reduces to the classical loop structure of bright solitons \cite{LIA2}; the periodic recurrence of the loop is gone.

The Kuznetsov-Ma breather can transform into the Peregrine rogue wave \cite{PRW} with double localization in the limit of $b\rightarrow \kappa_0$. The latter is also the limiting case of the Akhmediev breathers. All these breathers can induce loop-structure excitations on vortex filaments, as shown above and in Ref. \cite{AB,multibreather}.

One then wonder how to identify explicitly these vortex filaments induced by breathers, since each kind of vortex filament has a similar loop structure corresponding to the maximum curvature. This is the question of fundamental importance that has not been answered before. To do this, we introduce the following physical quantity---\textit{the integral of the relative quadratic curvature} of the form
\begin{eqnarray}
\Delta K=\int_{-\infty}^\infty \left[\kappa^2\left(s,t\right)-\kappa_0^2\left(s,t\right)\right]ds.\label{eqqc}
\end{eqnarray}
Eq. (\ref{eqqc}) corresponds to the \emph{effective energy} of breathers in optics \cite{NA1,NA2}. Namely, it coincides with the energy of breathers against plane wave, i.e., $\int_{-\infty}^\infty \left(\psi^2-\psi_0^2\right)ds$. For a quantum condensate fluid, Eq. (\ref{eqqc}) stands for the effective atom numbers \cite{quantumfluid}.
Generally, this is a quantity of physical importance which can be monitored effectively for localised nonlinear waves in experiments \cite{PN,quantumfluid}.
Here we highlight that Eq. (\ref{eqqc}) can be used for characterising the breather induced vortex filaments in quantized superfluid.

It is interesting to note that for the Kuznetsov-Ma breather (\ref{equ:KMB}), one obtains exactly $\Delta K=8\sqrt {b^2-\kappa_0^2}$, which indicates $\Delta K>0$; while for the Peregrine rogue wave and the Akhmediev breather,
we find that $\Delta K=0$ (see Appendix \ref{Appendix C}). This is the immanent reason why the loop structure induced by Kuznetsov-Ma breather exhibits periodic oscillation as $t$ increases, while the loop structure triggered by the Peregrine rogue wave and the Akhmediev breather appears only once during the time evolution.
On the other hand, the condition $\Delta K=0$ indicates that the resulting vortex filament starts from a uniform helical vortex structure.
This corresponds to the case of the Peregrine rogue wave and the Akhmediev breather. However, the uniform helical vortex structure will never appear for the vortex filament induced by the Kuznetsov-Ma breather.

\begin{figure}[htbp]
\centering
\includegraphics[width=82mm]{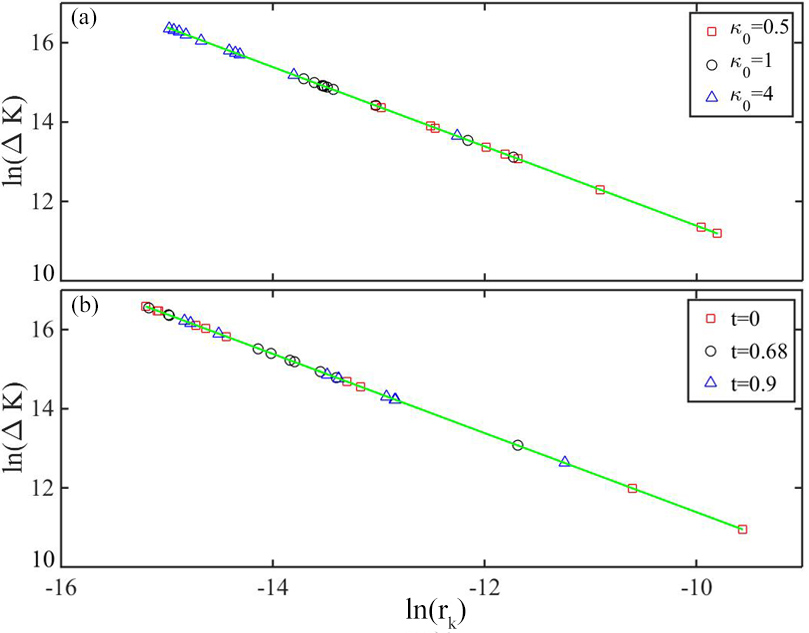}\\
\caption{Relations between $\Delta K(\kappa_0, t)$ and $r_k(\kappa_0, t)$ on logarithmic coordinates ($\ln{\Delta K}$, $\ln{r_k}$) (a) as $\kappa_0$ varies with fixed $t=-0.01$; (b) as $t$ varies with fixed $\kappa_0=1$. The solid lines are precise description of relation between $\ln{\Delta K}$ and $\ln{r_k}$ as $b\rightarrow \infty$. The values of the parameter $b$ are random numbers in the region $b\in[2\kappa_0, 30\kappa_0]$.
}\label{fig3}
\end{figure}

\begin{figure}[htbp]
\centering
\includegraphics[width=80mm]{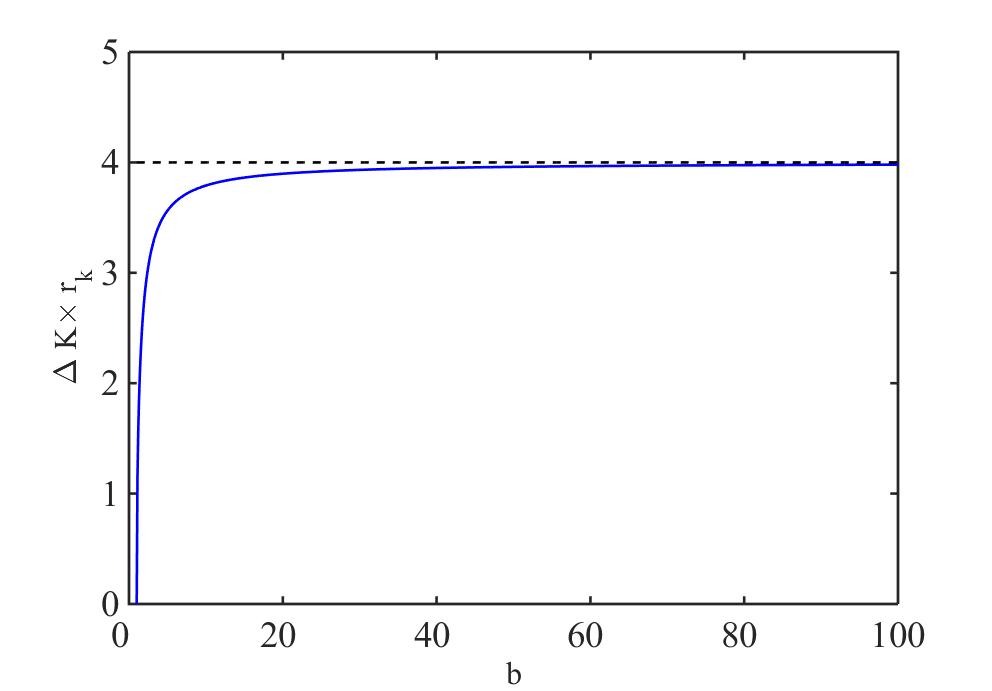}\\
\caption{Profile of $\Delta K\cdot r_k$, Eq. (\ref{eqlr2}) as $b$ increases. Other parameters are $\kappa_0=1$, $t=2n\pi/(\eta\beta)$ ($n$ is an integer) and $\beta=4\pi$.}\label{fig4}
\end{figure}

Let us take a closer look at Eq. (\ref{eqqc}) by considering the relation between $\Delta K$ and the Kuznetsov-Ma breather-induced loop structure. To this end, we define the characteristic size of the loop structure, $r_k$, which describes the minimum radius of the structure. 

Figure \ref{fig3} shows the relation between $\Delta K(\kappa_0, t)$ and $r(\kappa_0, t)$ on logarithmic coordinates with \textit{random values} of $b$ in the region $b\in[2\kappa_0,\infty]$. This parameter condition allows us to study the qualitative link between $\Delta K(\kappa_0, t)$ and $r_k(\kappa_0, t)$ from the vortex filaments induced by random Kuznetsov-Ma breathers (i.e., a series of Kuznetsov-Ma breathers with random period and amplitude).

For the fixed $t$ ($t=0$), we show the characteristics of $\ln(\Delta K)$ and $\ln(r_k)$ with increasing $\kappa_0$ in Fig. \ref{fig3}(a).
It is interesting that, despite the random values of $b$,
$\ln(\Delta K)$ decreases \textit{linearly} as $\ln(r_k)$ increases and the corresponding rates $\alpha$ are exactly consistent at a fixed time ($\alpha=-1$).
The similar linear relation also holds for the case with fixed $\kappa_0$ and $\tau_0$ and variational $t$, as shown in Fig. \ref{fig3}(b).

We then explain the linear relation above exactly.
We note that for the Kuznetsov-Ma-breather-induced loop structure, the minimum loop radius $r_k$ is inversely proportional to the maximum curvature $\kappa_{m}$, i.e., $$r_k=\frac{1}{\kappa_{m}}.$$
It is given explicitly by Eq. (\ref{equ:11}) at $\xi=0$:
\begin{eqnarray}
\label{equ:13}
r_k=\left[\left(\kappa_0+\frac{2\chi^2\cos(\eta\beta t)}{n_2}\right)^2+\frac{4\eta^2\sin^2(\eta\beta t)}{\left(n_2\right)^2}\right]^{-1/2}
\end{eqnarray}
with $n_2=\kappa_0\cos\left(\eta\beta t\right)-b$. Here $r_k$ is the function of $b$, $\kappa_0$ and $t$. Thus the accurate description of $\Delta K \cdot r_k$ reads
\begin{equation}\label{eqlr2}
\Delta K\cdot r_k=\frac{8\sqrt {b^2-\kappa_0^2}}{\left[\left(\kappa_0+\frac{2\chi^2\cos(\eta\beta t)}{n_2}\right)^2+\frac{4\eta^2\sin^2(\eta\beta t)}{\left(n_2\right)^2}\right]^{1/2}}.
\end{equation}
Clearly, for the case of Kuznetsov-Ma breather, $\Delta K\cdot r_k\neq0$; while for case of the Peregrine rogue wave and Akhmediev breather, $\Delta K\cdot r_k=0$, since $\Delta K=0$.

We show the profile of Eq. (\ref{eqlr2}) as $b$ increases in Fig. \ref{fig4}.
One can see that $\Delta K\cdot r_k$ increases monotonously with increasing $b$.
Remarkably, as $b\rightarrow \infty$, we find $\Delta K\cdot r_k\rightarrow 4$.
Indeed, one can readily obtain a simple relation from Eq. (\ref{eqlr2}) as $b\rightarrow \infty$
\begin{equation}\label{eqlre}
\Delta K\cdot r_k=4,
\end{equation}
Namely,
\begin{equation}
\ln\Delta K=-\ln r_k+\ln4,
\end{equation}
on logarithmic coordinates.

We show the linear relation by the solid lines in Fig. \ref{fig3}.
Observably, the numerical results are in good agreement with the analytical relation (\ref{eqlre}) (solid line).
Physically, the Kuznetsov-Ma breather in the region $b\in[2\kappa_0,\infty]$ can be approximatively described by the \emph{linear interference} between a bright soliton and a plane wave \cite{Mechanism of KMB}. As $b\rightarrow \infty$ ($b\gg\kappa_0$), i.e., the amplitude of the bright soliton is much bigger than that of the plane wave, the plane wave can be neglected.
As a result, the effective energy $\Delta K$ is quadruple of the amplitude of the remaining bright soliton, which directly leads to Eq. (\ref{eqlre}).

\begin{figure*}[htbp]
  \centering
  \includegraphics[height=95mm,width=115mm]{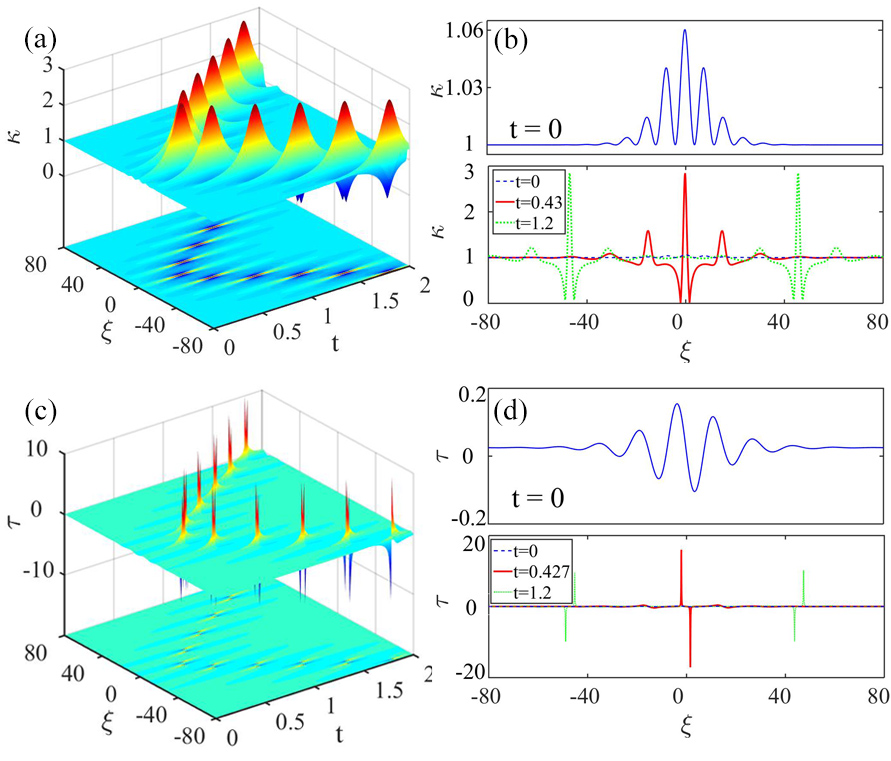}\\
  \caption{Temporal evolutions of curvature $\kappa\left(\xi,t\right)$ (a) and torsion $\tau\left(\xi,t\right)$ (c) corresponding to a super-regular breather, see Eq. (\ref{eqsr}) in Appendix \ref{Appendix B}. (b) and (d) are variations of $\kappa\left(\xi,t\right)$ and $\tau\left(\xi,t\right)$ at different times.
  Other parameters are $\kappa_0=1$, $\tau_0=0.05$, $R=1.1$ and $\phi=\pi/8$.}\label{fig5}
\end{figure*}

\section{super-regular breather induced loop pair and symmetry breaking}
Let us then consider the vortex filament induced by the super-regular breather. The latter, which recently serves as the exact MI scenario excited from localised multi-peak perturbations, is formed by the nonlinear superposition of two quasi-Akhmediev breathers \cite{SR1,SR3,SR4,SR5,SR6}. The exact solution of super-regular breather with $\tau_0=0$ is first provided in Ref. \cite{SR1}. However, the general solution with $\tau_0\neq0$ in the infinite NLSE is presented recently in Ref. \cite{SR3}. By using the transformation above and the super-regular solution in Ref. \cite{SR3}, the corresponding properties of vortex filaments can be achieved effectively. Here we omit the tedious explicit expression but show the important and compact results.
At the first step, the integral of the relative quadratic curvature of super-regular breather can be obtained explicitly from the exact solution in Appendix \ref{Appendix B}. It reads,
\begin{equation}\label{eqqcsr}
\Delta K=16\kappa_0\left[\varepsilon~\cos\phi+\pi~\sin\phi~{\rm csch}\left(\frac{\pi~\sin\phi}{\varepsilon~\cos\phi}\right)\right],
\end{equation}
where $\varepsilon=R-1$ and $\varepsilon\ll1$. $R(>1)$ and $\phi[\in(-\pi/2,\pi/2)]$ are two real parameters that denote respectively
the radius and angle in polar coordinates (see Appendix \ref{Appendix B}).
Physically, $R$ (or $\varepsilon$) and $\phi$ are two important parameters that describe directly the amplitude and period of the super-regular breathers. It is therefore crucial to study the property of vortex filaments induced by super-regular breathers by the choice of parameters $R$ and $\phi$.

Just as the case of Kuznetsov-Ma breather, Eq. (\ref{eqqcsr}) is also greater than zero, i.e., $\Delta K>0$. This indicates that the super-regular breather also admits long-time dynamics which is different from the Peregrine rogue wave and Akhmediev breather. Unlike the case of Kuznetsov-Ma breather, the evolutions of curvature and torsion of super-regular breather exhibit remarkably different characteristics. This stems from that the super-regular breather possesses a localised multi-peak perturbation rather than a localised single-peak perturbation.

Figure \ref{fig5} shows the variation of curvature and torsion induced by super-regular breather with the initial parameters $\kappa_0=1$, $R=1.1$, and $\phi=\pi/8$. As can be seen from Fig. \ref{fig5}(a) that the curvature of a super-regular breather triggered from a localised multi-peak perturbation at $t=0$ [see Fig. \ref{fig5}(b)] increases gradually due to the exponential amplification of the MI at the linear stage. It reaches its maximum at $t=0.43$ and then splits into two quasi-Akhmediev breathers propagating along different directions during the nonlinear stage of MI. The corresponding torsion also suffers singular behavior starting from the maximum curvature point $t=0.43$ [see Fig. \ref{fig5}(c)]. Interestingly, the nonlinear propagation stage always holds the singular torsion at the maximum curvature point as $t>0.43$ [see Fig. \ref{fig5}(d)].

\begin{figure}[htb]
  \centering
  \includegraphics[width=85mm]{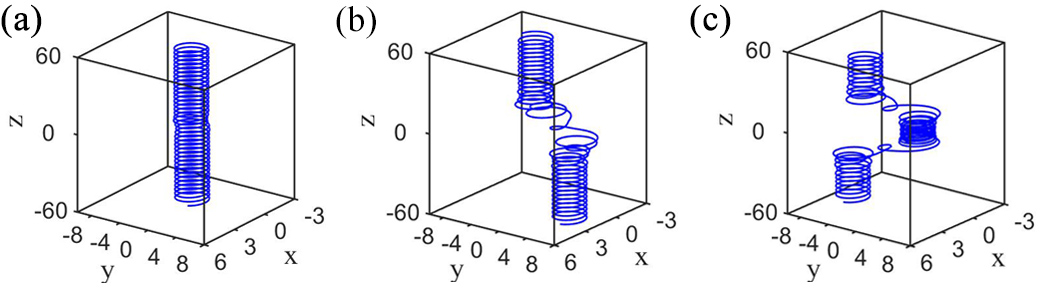}\\
  \caption{Configuration of vortex filaments of super-regular breathers at different time (a) $t=0$, (b) $t=0.43$ and (c) $t=1.2$. Other parameters are the same as in Fig. \ref{fig4}.}\label{fig6}
\end{figure}

\begin{figure}[htb]
\centering
\includegraphics[width=85mm]{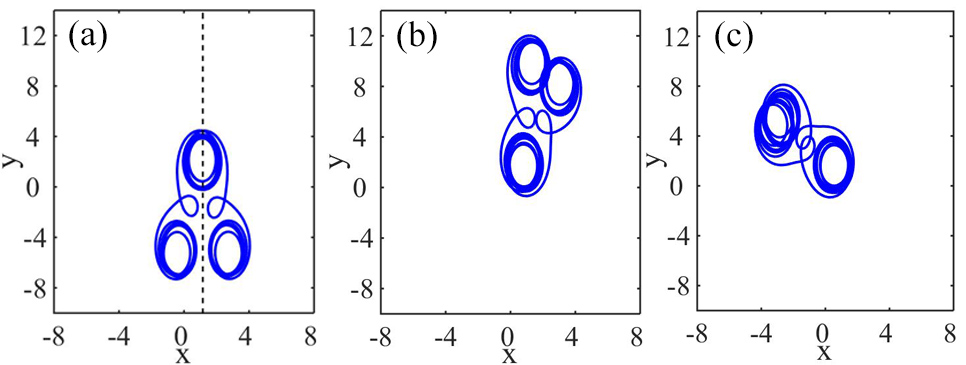}\\
\caption{Top view of configuration of vortex filaments of super-regular breathers at a fixed time $t=1.2$ as $\tau_0$ increases. (a) $\tau_0=0.01$, (b) $\tau_0=0.17$ and (c) $\tau_0=0.24$. One can see clearly the reflection symmetry breaking of the loop pair with non-zero $\tau_0$.
 Other parameters are $\kappa_0=1$, $R=1.1$ and $\phi=\pi/8$.}\label{fig7}
\end{figure}

Figure \ref{fig6} displays the corresponding vortex structure at $t=0$, $t=0.43$, and $t=1.2$, respectively. One can see clearly that the vortex filament emerges from a perturbed helical vortex at $t=0$ and then exhibits a remarkable loop structure at $t=0.43$.
This is the linear MI stage that corresponds to one loop excitation. Interestingly, once the vortex filament evolves into the nonlinear stage, the loop structure splits into a \emph{loop pair} which corresponds to the two quasi-Akhmediev breathers propagation with different group velocities.

In particular, we find that the loop pair induced by super-regular breather at the nonlinear stage shows an interesting \textit{reflection symmetry breaking}, as shown in Fig. \ref{fig7}. We find that this remarkable feature comes from the asymmetry of the group velocities of the two quasi-Akhmediev breathers. Indeed, the group velocities of the super-regular breather are given by [see Eq. (\ref{eqsr1}) in Appendix \ref{Appendix B}]
\begin{equation}                                                                                                                                                                                                                                                                 V_{g1}=2\beta\tau_0+d,~~V_{g2}=2\beta\tau_0-d,
\end{equation}
where $d=\beta\kappa_0\frac{\left(R^4+1\right)}{R^3-R}\sin{\phi}$. Clearly, due to $\tau_0\neq0$, the absolute values of this two group velocities are always unequal.
Once $\kappa_0$, $\varepsilon$, and $\phi$ are fixed, the degree of the asymmetry is proportional to the value of $|\tau_0|$.

Figure \ref{fig7} shows the corresponding vortex structures induced by super-regular breather as $\tau_0$ increases.
We see that as $\tau_0\rightarrow0$ the resulting loop pair exhibits quasi-reflection symmetry [Fig. \ref{fig7}(a)], while
the reflection symmetry of the loop pair breaks greatly with increasing $\tau_0$ [Figs. \ref{fig7}(b) and \ref{fig7}(c)].

It is very interesting to note that, despite the broken reflection symmetry as $\tau_0\neq0$, the growth rate of modulation instability driven by
the super-regular breather does not depend on $\tau_0$. Namely, this growth rate is only associated with the absolute difference of the group velocities, $G=\eta_r|V_{g1}-V_{g2}|$ with $\eta_r=\frac{a}{2}\left(R-1/R\right)\cos{\phi}$, as shown in Ref. \cite{SR3}.
This result is physically important because that although the super-regular breather induced vortex structures can exhibit different loop pairs with symmetry breaking, the inherent MI property can remain invariable.

Finally, we consider the relation between $\Delta K$ and characteristic size $r_s$ of the super-regular breather induced vortex structures. Similar to the case of Kuznetsov-Ma breather, we define characteristic size $r_s$ as the minimum radius of the super-regular-breather induced vortex structure throughout the whole evolution.
Thus, the characteristic size $r_s$, which is also inversely
proportional to the maximum curvature $\kappa_{ms}$ (i.e., $r_s=1/\kappa_{ms}$), is given by
\begin{equation}\label{rs}
r_s=\left[\kappa_0+\kappa_0\left(1+\varepsilon+\frac{1}{1+\varepsilon}\right)\cos\phi\right]^{-1},
\end{equation}
where $\varepsilon=R-1$ is a small value ($\varepsilon\ll1$) defined above.

Collecting Eq. (\ref{eqqcsr}) and Eq. (\ref{rs}), we obtain the explicit expression of $\Delta K\cdot r_s$ by omitting the high-order term $O(\varepsilon^2)$. It reads
\begin{equation}\label{Krs2}
\Delta K\cdot r_s=\alpha_s\varepsilon,
\end{equation}
where $\alpha_s=16\cos\phi/\left(1+2\cos\phi\right)$.

In contrast to the case of Kuznetsov-Ma breather, Eq. (\ref{eqlr2}), where only one parameter $b$ can be modulated
when the plane wave parameters ($\kappa_0$, $\tau_0$) and the structural parameter $\beta$ are fixed,
Eq. (\ref{Krs2}) has two free physical parameters ($\varepsilon$ and $\phi$). But even so, we highlight that linear relations can also hold for the case of super-regular breather induced vortex structures.

\begin{figure}[htb]
\centering
\includegraphics[width=82mm]{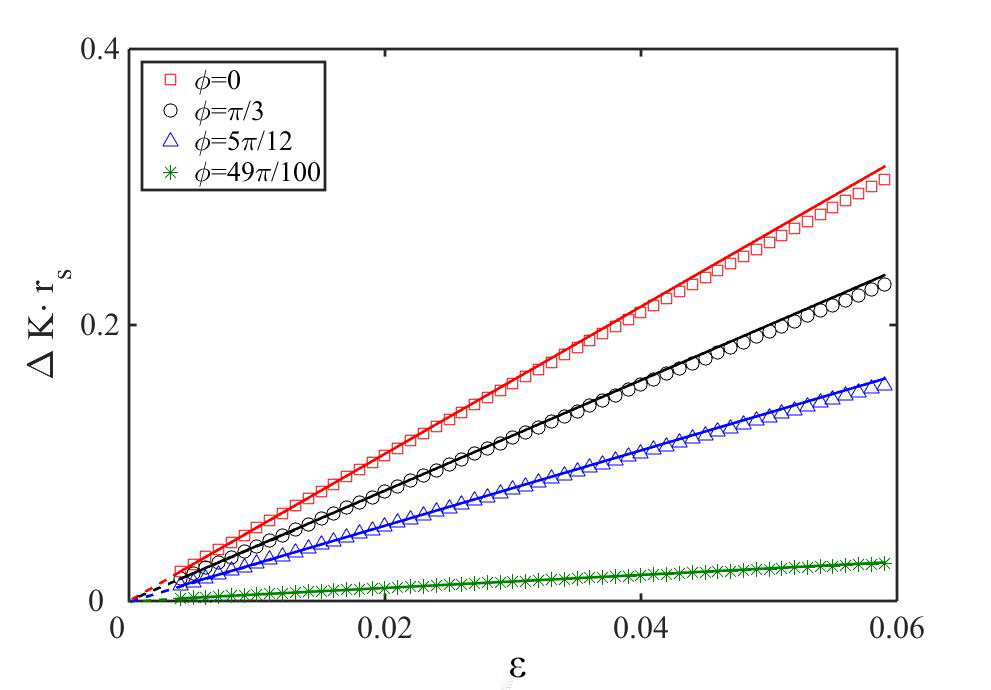}\\
\caption{Relations between $\Delta K\cdot r_s$ and $\varepsilon$ of the vortex filament induced by super-regular breather as $\phi$ varies.
 Other parameters are $\kappa_0=1$. The discrete points are obtained with the high-order term $O(\varepsilon^2)$ considered, while the colored lines retain the first-order term only. }\label{fig8}
\end{figure}
\begin{figure}[htb]
\centering
\includegraphics[width=82mm]{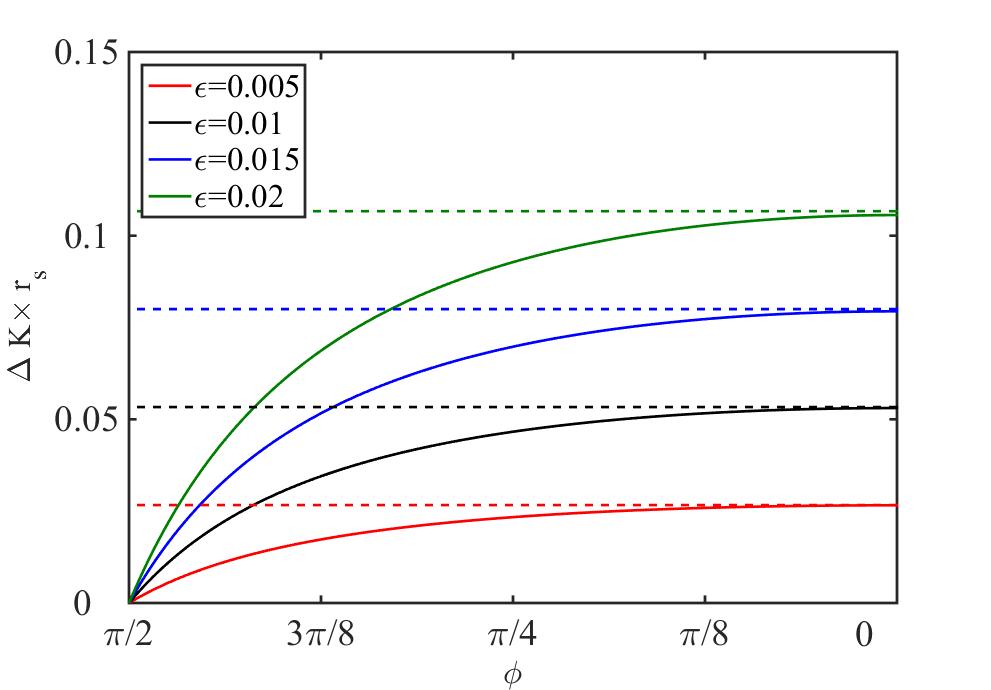}\\
\caption{Relations between $\Delta K\cdot r_s$ and $\phi$ of the vortex filament induced by super-regular breather as $\varepsilon$ varies.
 Other parameters are $\kappa_0=1$.}\label{fig9}
\end{figure}

Figure \ref{fig8} shows the characteristics of $\Delta K\cdot r_s$ as $\varepsilon$ increases with different values of $\phi$.
In particular, we compare the results obtained from the approximate expression Eq. (\ref{Krs2}) (the solid lines) and the exact expression (the dotted lines), respectively.
One can see that for each fixed $\phi$, $\Delta K\cdot r_s$ shows a linear relation with $\varepsilon$. The corresponding rate $\alpha_s$ decreases in the range of $[48,0]$ as $\phi$ increases from $0$ to $\pi/2$.

Figure \ref{fig9} shows the characteristics of $\Delta K\cdot r_s$ as $\phi$ decreases with different values of $\varepsilon$.
For each case with fixed $\varepsilon$, $\Delta K\cdot r_k$ increases monotonously with decreasing $\phi$.
As $\phi\rightarrow0$, one obtains that $\Delta K\cdot r_k\rightarrow16\varepsilon/3$. This is similar with the case of Kuznetsov-Ma breather induced vortex structures shown in Fig. \ref{fig4}.
Physically, as $\phi\rightarrow0$, the super-regular breather
transforms itself into two colliding Kuznetsov-Ma breathers, so that the similar linear relation can be maintained.

\section{CONCLUSION}
In summary, we have investigated the superfluid vortex filaments induced by Kuznetsov-Ma breather and super-regular breather, which admit localised perturbations. We have shown that the loop structure induced by Kuznetsov-Ma breather emerges periodically as time increases, while the loop structure triggered by super-regular breather---the loop pair---exhibits striking symmetry breaking due to the broken reflection symmetry of the group velocities of super-regular breather.
In particular, we have characterized and identified explicitly these loop excitations by introducing the integral of the relative quadratic curvature, which corresponds to the effective energy of breathers. Although the nature of nonlinearity, it is demonstrated that this physical quantity shows a linear correlation with the loop size.

\section*{ACKNOWLEDGEMENTS}
This work has been supported by the National Natural Science Foundation of China (NSFC) (Grant Nos. 11705145, 11875220, 11434013, and 11425522), Natural Science Basic Research Plan in Shaanxi Province of China (Grant No. 2018JQ1003), and the Major Basic Research Program of Natural Science of
Shaanxi Province (Grant Nos. 2017KCT-12, 2017ZDJC-32).

\begin{appendix}

\section{THE POSITION VECTOR OF THE VORTEX FILAMENT}\label{Appendix A}
We give the explicit expression of the position vector of the vortex filament by integrating the Frenet-Serret equations given in \cite{SFe} and the exact expression is formulated explicitly in Ref. \cite{model2}, which should be represented as
\begin{eqnarray}
\label{eqqa1}
\nonumber
\mathbf{r}\left(s,t\right)&&=
\begin{bmatrix}
x\left(s,t\right)\\
y\left(s,t\right)\\
z\left(s,t\right)
\end{bmatrix}\\\nonumber
&&=\begin{bmatrix}
x_0\left(t\right)+\displaystyle\sum_{k=1}^3c_{k1}\left(t\right)\int_0^sM_{k}\left(\sigma,t\right)\,d\sigma\\
y_0\left(t\right)+\displaystyle\sum_{k=1}^3c_{k2}\left(t\right)\int_0^sM_{k}\left(\sigma,t\right)\,d\sigma\\
z_0\left(t\right)+\displaystyle\sum_{k=1}^3c_{k3}\left(t\right)\int_0^sM_{k}\left(\sigma,t\right)\,d\sigma
\end{bmatrix}.
\end{eqnarray}
Here, $x_0\left(t\right)$, $y_0\left(t\right)$, $z_0\left(t\right)$ are constants with respect to the initial position of vortex structures. $M_{k}\left(k=1,2,3\right)$ are
\begin{eqnarray}
M_{1}=\frac{\gamma^2+\alpha^2\cos{\lambda}}{\lambda^2},M_{2}=\frac{\alpha\sin{\lambda}}{\lambda},
M_{3}=\frac{\alpha\gamma\left(1-\cos{\lambda}\right)}{\lambda^2}\nonumber,
\end{eqnarray}
where $\alpha=\int_0^s\kappa\left(\sigma,t\right)\,d\sigma$, $\gamma=\int_0^s\tau\left(\sigma,t\right)\,d\sigma$, and\\ $\lambda=\sqrt{\left(\int_0^s\kappa\left(\sigma,t\right)\,d\sigma\right)^2+\left(\int_0^s\tau\left(\sigma,t\right)\,d\sigma\right)^2}$.

\section{EXPLICIT EXPRESSIONs OF $\Delta K$ OF AKHMEDIEV BREATHER AND PEREGRINE ROGUE WAVE}\label{Appendix C}
$\Delta K$, \textit{the integral of the relative quadratic curvature}, is expressed explicitly in the form
\begin{equation}
\label{eqqc1}
\Delta K=\int_{-\infty}^\infty \left[\kappa^2\left(s,t\right)-\kappa_0^2\left(s,t\right)\right]ds.
\end{equation}
Here, $\kappa\left(s,t\right)$ and $\kappa_0\left(s,t\right)$ represent the curvature distribution of the vortex filament and the curvature of the background uniform helical vortex corresponding to plane wave $\psi_0$ (\ref{equ:PWB}) respectively.

As mentioned in Sec \ref{Sec3}, in addition to the Kuznetsov-Ma breather, plane wave (\ref{equ:PWB}) also admits other breathing waves, including the Akhmediev breather and the Peregrine rogue wave. As comparison, we show here $\Delta K$ for the vortex filaments induced by Akhmediev breather and Peregrine rogue wave.

We first consider the Akhmediev breather that exhibits the explicit description for the MI emerging from periodic perturbations. Its exact expressions is given by
\begin{equation}\label{eqqc2}                                                                                                                                                                                                                                                              \psi_A(s,t)=\left[1-2\frac{\chi_1^2\cosh\left(\eta_1\beta t\right)+i\eta_1 \sinh\left(\eta_1\beta t\right)}{\kappa_0^2~\cos\left(\eta_1\beta t\right)-\kappa_0b~\cosh\left(\chi_1\xi\right)}\right]\psi_0,
\end{equation}
where $\chi_1=\sqrt{\kappa_0^2-b^2}$ with $b<\kappa_0$, $\eta_1=b\chi_1$, and $\xi=s-2\tau_0\beta t$. The corresponding exact expression of the curvature is given by
\begin{equation}
\label{eqqc3}
\kappa_A=\left[\left(\kappa_0-\frac{2\chi_1^2\cosh(\eta_1\beta t)}{A}\right)^2+\frac{4\eta_1^2\sinh^2(\eta_1\beta t)}{A^2}\right]^{1/2}
\end{equation}
with $A=\kappa_0\cosh\left(\eta_1\beta t\right)-b\cos\left(\chi_1\xi\right)$.
A substitution of Eq. (\ref{eqqc3}) into  Eq. (\ref{eqqc1}) yields $\Delta K_A=0$.

We then consider the Peregrine rogue wave with double localization.
The latter corresponds to the limiting case of Eq. (\ref{eqqc2}) as $b\rightarrow \kappa_0$.
Its exact expression is given by
\begin{equation}
\psi_P\left(s,t\right)=\left[1-\frac{4i\kappa_0^2\beta t+4}{1+\kappa_0^4\beta^2 t^2+\kappa_0^2\left(s-2\beta\tau_0 t\right)^2}\right]\psi_0,
\end{equation}
whose curvature is in the form of
\begin{equation}
\label{eqqc4}
\kappa_P=\kappa_0\sqrt{\frac{16\kappa_0^4\beta^2 t^2}{a^2}+\left(1-\frac{4}{a}\right)^2}
\end{equation}
with $a=1+\kappa_0^4\beta^2t^2+\kappa_0^2\left(s-2\beta\tau_0 t\right)^2$.
By calculating Eq. (\ref{eqqc1}), we demonstrate also that  $\Delta K_P=0$.

As a result, both the Akhmediev breather and the Peregrine rogue wave share the vanishing $\Delta K$, which indicates that the corresponding vortex filaments start from a uniform helical vortex structure.

\section{EXPLICIT EXPRESSION OF SUPER-REGULAR BREATHER}\label{Appendix B}
The explicit expression of super-regular breather for Eq. (\ref{equ:NLSE}) is given by the Darboux transformation \cite{SR3}, where the spectral parameter $\lambda$ is parameterized by the Jukowsky transform \cite{SR1} as follows:
\begin{eqnarray}\label{eqlm}
\lambda=i\frac{\kappa_0}{2}\left(\Delta+\frac{1}{\Delta}\right)-\frac{\tau_0}{2},~\Delta=Re^{i\phi}.
\end{eqnarray}
Here, $R$ and $\phi$ define the location of the spectral parameter $\lambda$ in the polar coordinates. They represent radius and angle respectively  in the region $R>1$ and $\phi\in\left(-\pi/2,\pi/2\right)$.
For $\tau_0=0$, Eq (\ref{eqlm}) reduces to the spectral parameter used in Ref. \cite{SR1}.
With different values of $R$ and $\phi$, the resulting exact solution can describe different breather dynamics \cite{SR1}. A more general phase diagram of breathers has been obtained recently in Ref. \cite{CLIP}.
Here we consider the super-regular breather formed by two quasi-Akhmediev breathers with $R_1=R_2=R=1+\varepsilon$ ($\varepsilon\ll1$), $\phi_1=-\phi_2=\phi$. Its explicit expression of the solution is in the form:
\begin{eqnarray}\label{eqsr}
\psi\left(s,t\right)=\psi_0\left[1-4\rho\varrho\frac{\left(i\varrho-\rho\right)\Xi_1+\left(i\varrho+\rho\right)\Xi_2}{\kappa_0\left(\rho^2\Xi_3+\varrho^2\Xi_4\right)}\right].
\end{eqnarray}
Here \begin{eqnarray}
&&\varrho=\frac{\kappa_0}{2}\left(R-\frac{1}{R}\right)\sin\phi, ~\rho=\frac{\kappa_0}{2}\left(R+\frac{1}{R}\right)\cos\phi\nonumber\\
&&\Xi_1=\varphi_{21}\phi_{11}+\varphi_{22}\phi_{21}, ~~~\Xi_2=\varphi_{11}\phi_{21}+\varphi_{21}\phi_{22},\nonumber\\
&&\Xi_3=\varphi_{11}\phi_{22}-\varphi_{21}\phi_{12}-\varphi_{12}\phi_{21}+\varphi_{22}\phi_{11},\nonumber\\
&&\Xi_4=\left(\varphi_{11}+\varphi_{22}\right)\left(\phi_{11}+\phi_{22}\right),\nonumber
\end{eqnarray}
with
\begin{eqnarray}
&&\phi_{jj}=\cosh\left(\Theta_2\mp i\psi\right)-\cos\left(\Phi_2\mp\phi\right),\nonumber\\
&&\varphi_{jj}=\cosh\left(\Theta_1\mp i\psi\right)-\cos\left(\Phi_1\mp\phi\right),\nonumber\\
&&\phi_{j3-j}=\pm i\cosh\left(\Theta_2\mp i\phi\right)-\cos\left(\Phi_2\mp\theta\right),\nonumber\\
&&\varphi_{j3-j}=\pm i\cosh\left(\Theta_1\mp i\phi\right)-\cos\left(\Phi_1\mp\theta\right),\nonumber
\end{eqnarray}
where $\theta=\arctan\left[\left(1-iR^2\right)/\left(1+R^2\right)\right]$. $\Theta_j$ and $\phi_j$ are related with group and phase velocities respectively, which is in the form of
\begin{eqnarray}
\Theta_j=2\eta_r\left(s-V_{gj}t\right),~\phi_j=2\eta_{ij}\left(s-V_{pj}t\right)
\end{eqnarray}
where
\begin{eqnarray}
&&\eta_{i1}=-\eta_{i1}=\frac{\kappa}{2}\left(R+\frac{1}{R}\right)\sin\phi,\nonumber\\ &&\eta_r=\frac{\kappa}{2}\left(R-\frac{1}{R}\right)\cos\phi,\nonumber\\ &&V_{p1}=2\beta\tau_0-d_1,~V_{p2}=2\beta\tau_0+d_2,\nonumber\\
&&V_{g1}=2\beta\tau_0+d,~V_{g2}=2\beta\tau_0-d.\label{eqsr1}
\end{eqnarray}
with $d_1=\beta\kappa_0\left(R-\frac{1}{R}\right)\frac{\cos\left(2\phi\right)}{\sin\phi}$, $d_2=\beta\kappa_0\frac{\left(R-\frac{1}{R}\right)}{\sin\phi}$ and $d=\beta\kappa_0\frac{\left(R^4+1\right)}{R^3-R}\sin{\phi}$.
The initial state of the super-regular breather can be extracted from the above solution at $t=0$.
It reads, as $\varepsilon\rightarrow0$,
\begin{eqnarray}\label{eqsrin}
\psi\left(s,0\right)=\psi_0 \left(1-i\frac{4\varepsilon \cos\phi \cos\left(\kappa_0 s \sin\phi\right)}{\cosh\left(\kappa_0\varepsilon s \cos\phi\right)}\right).
\end{eqnarray}
Note that for a given plane wave background (\ref{equ:PWB}) (i.e., $\kappa_0$ and $\tau_0$ are fixed), $R$ and $\phi$ determine the amplitude and period of breathers. In particular, $R$ and $\phi$ effect the profile of the initial state (\ref{eqsrin}) of super-regular breathers.

The integral of the relative quadratic curvature of super-regular breather $\Delta K$ [see Eq. (\ref{eqqcsr})] is obtained explicitly from the initial state (\ref{eqsrin}), since the
NLSE shares the same $\Delta K$ at different times.

\end{appendix}


\begin{thebibliography}{}
\bibitem{quantumfluid} C. F. Barenghi and N. G. Parker, \textit{A primer on quantum fluids} (Springer, 2016).
\bibitem{superfluid1} I. Carusotto and C. Ciuti, \href{https://journals.aps.org/rmp/abstract/10.1103/RevModPhys.85.299}{Rev. Mod. Phys. \textbf{85}, 299 (2013)}.
\bibitem{vortex reconnection} S. Zuccher, M. Caliari, A. Baggaley, and C. Barenghi, \href{https://aip.scitation.org/doi/abs/10.1063/1.4772198}{Phys. Fluids \textbf{24}, 125108 (2012)}.
\bibitem{vortices1} S. K. Nemirovskii, \href{https://www.sciencedirect.com/science/article/pii/S037015731200350X}{Phys. Rep. \textbf{524}, 85 (2013)}.
\bibitem{BS} P. G. Saffman, \textit{Vortex Dynamics} (Cambridge University Press, Cambridge, 1992).
\bibitem{vortex tangles1} W. Gilpin, V. N. Prakash, M. Prakash, \href{https://www.nature.com/articles/nphys3981}{Nat. Phys. \textbf{13}, 380 (2017)}.
\bibitem{LIA2} H. Hasimoto, \href{https://www.cambridge.org/core/journals/journal-of-fluid-mechanics/article/soliton-on-a-vortex-filament/1A2FF58DD1B8DE20509866A4713531F3}{J. Fluid Mech \textbf{51}, 477 (1972)}.
\bibitem{AB} H. Salman, \href{https://journals.aps.org/prl/abstract/10.1103/PhysRevLett.111.165301}{Phys. Rev. Lett \textbf{111}, 165301 (2013)}.
\bibitem{multibreather} H. Salman, \href{http://iopscience.iop.org/article/10.1088/1742-6596/544/1/012005/meta}{J. Phys.: Conf. Ser. \textbf{544}, 012005 (2014)}.
\bibitem{excitations1} B. K. Shivamoggi, \href{https://journals.aps.org/prb/abstract/10.1103/PhysRevB.84.012506}{Phys. Rev. B \textbf{84}, 012506 (2011)}.
\bibitem{excitations2} R. A. Van Gorder, \href{https://journals.aps.org/pre/abstract/10.1103/PhysRevE.93.052208}{Phys. Rev. E \textbf{93}, 052208 (2016)}; \href{https://journals.aps.org/pre/abstract/10.1103/PhysRevE.91.053201}{Phys. Rev. E 91, 053201 (2015)}; R. Shah, R. A. Van Gorder, \href{https://journals.aps.org/pre/abstract/10.1103/PhysRevE.93.032218}{Phys. Rev. E 93, 032218 (2016)}.
\bibitem{KW1} W.T. Kelvin, Vibrations of a columnar vortex. Phil. Mag. Ser. 5, 155-168 (1880).
\bibitem{KW2} E. Kozik and B. Svistunov, \href{https://journals.aps.org/prl/abstract/10.1103/PhysRevLett.92.035301}{Phys. Rev. Lett. 92, 035301 (2004)}.
\bibitem{Book97} N. Akhmediev and A. Ankiewicz, \emph{Solitons: Nolinear Pulses and Beams} (Chapman and Hall, London, 1997).
\bibitem{O1a} J. M. Dudley, F. Dias, M. Erkintalo, and G. Genty,
\href{https://www.nature.com/articles/nphoton.2014.220}{Nat. Photonics \textbf{8}, 755 (2014)}.
\bibitem{SR1} V. E. Zakharov and A. A. Gelash, \href{https://journals.aps.org/prl/abstract/10.1103/PhysRevLett.111.054101}{Phys. Rev. Lett. \textbf{111}, 054101 (2013)}; A. A. Gelash and V. E. Zakharov, \href{http://iopscience.iop.org/article/10.1088/0951-7715/27/4/R1/meta}{Nonlinearity \textbf{27}, R1 (2014)}; B. Kibler, A. Chabchoub, A. Gelash, N. Akhmediev, and V. E. Zakharov,
\href{https://journals.aps.org/prx/abstract/10.1103/PhysRevX.5.041026}{Phys. Rev. X \textbf{5}, 041026 (2015)}.
\bibitem{OE} J. M. Dudley, G. Genty, F. Dias, B. Kibler, and N. Akhmediev, \href{https://www.osapublishing.org/oe/abstract.cfm?uri=oe-17-24-21497}{Opt. Express \textbf{17}, 21497 (2009)}.
\bibitem{MI1} L. C. Zhao and L. M. Ling, \href{https://www.osapublishing.org/josab/abstract.cfm?uri=josab-33-5-850}{J. Opt. Soc. \textbf{33}, 850 (2016)}.
\bibitem{Extreme09} N. Akhmediev, J. Soto-Crespo, and A. Ankiewicz,
\href{https://www.sciencedirect.com/science/article/pii/S0375960109004939}{Phys. Lett. A \textbf{373}, 2137 (2009)}; N. Akhmediev, J. Soto-Crespo, and A. Ankiewicz,
\href{https://journals.aps.org/pra/abstract/10.1103/PhysRevA.80.043818}{Phys. Rev. A \textbf{80}, 043818 (2009)}; N. Akhmediev, A. Ankiewicz, and M. Taki,
\href{https://www.sciencedirect.com/science/article/pii/S0375960108017945}{Phys. Lett. A \textbf{373}, 675 (2009)}.

\bibitem{W1b} M. Onorato, S. Residori, U. Bortolozzo, A. Montina, and F. T. Arecchi,
\href{https://www.sciencedirect.com/science/article/pii/S0370157313000963}{Phys. Rep. \textbf{528}, 47 (2013)}.
\bibitem{O1b} N. Akhmediev, {\it et al},
\href{http://iopscience.iop.org/article/10.1088/2040-8978/18/6/063001/meta}{J. Optics \textbf{18}, 063001 (2016)}.
\bibitem{Onorato} M. Onorato, S. Residori, and F. Baronio, \textit{Rogue and Shock Waves in
Nonlinear Dispersive Media} (Springer, 2016).
\bibitem{Wabnitz} S. Wabnitz, \emph{Nonlinear Guided Wave Optics: a testbed for extreme waves} (IOP Publ., Bristol, UK, 2017).
\bibitem{ABbreather} N. Akhmediev and V. I. Korneev, Theor. Math. Phys. \textbf{69}, 1089 (1986); N. Akhmediev, \href{https://www.nature.com/articles/35095154}{Nature \textbf{413}, 267 (2001)}.
\bibitem{KMB} E. Kuznetsov, Sov. Phys. Dokl. \textbf{22}, 507 (1977); Y. C. Ma, Stud. Appl. Math. \textbf{60}, 43 (1979).
\bibitem{Mechanism of KMB} L. C. Zhao, L. M. Ling, and Z. Y. Yang, \href{https://journals.aps.org/pre/abstract/10.1103/PhysRevE.97.022218}{Phys. Rev. E \textbf{97}, 022218 (2018)}.
\bibitem{SR3} C. Liu, Z. Y. Yang, and W. L. Yang, \href{https://aip.scitation.org/doi/10.1063/1.5025632}{Chaos, \textbf{28}, 083110 (2018)}.
\bibitem{SR4} Y. Ren, C. Liu, Z. Y. Yang, and W. L. Yang, \href{https://journals.aps.org/pre/abstract/10.1103/PhysRevE.98.062223}{Phys. Rev. E \textbf{98}, 062223 (2018)}.
\bibitem{LIA1} R. J. Arms, and F. R. Hama, \href{https://aip.scitation.org/doi/abs/10.1063/1.1761268}{The Physics of fluids \textbf{8}, 553 (1965)}.
\bibitem{liq3} L. S. Da Rios, Rend. Circ. Mat. Palermo \textbf{22}, 117 (1906).
\bibitem{liq4} R. Betchov, \href{https://www.cambridge.org/core/journals/journal-of-fluid-mechanics/article/on-the-curvature-and-torsion-of-an-isolated-vortex-filament/28809E8B5A4890DEE004F4F161ECD893}{J. Fluid Mech \textbf{22}, 471 (1965)}.
\bibitem{SFe} L. M. Pismen, \textit{Vortices in Nonlinear Fields} (Clarendon, Oxford, 1999).
\bibitem{model2} R. Shah, \textit{Rogue waves on a vortex filament} (Oxford, 2015).
\bibitem{PRW} D. H. Peregrine, \href{https://www.cambridge.org/core/journals/anziam-journal/article/water-waves-nonlinear-schrodinger-equations-and-their-solutions/D87F5416C657F3B5C35AE96DF9F73DD0}{J. Aust. Math. Soc. Ser. B, Appl. Math. \textbf{25}, 16 (1983)}.
\bibitem{NA1} N. Akhmediev and A. Ankiewicz, (Eds.), \emph{Dissipative solitons}, Lect. Notes Phys. 661 (Springer, Berlin Heidelberg 2005).
\bibitem{NA2} N. Akhmediev and A. Ankiewicz, (Eds.), \emph{Dissipative Solitons: From Optics to Biology and Medicine}, Lect. Notes Phys. 751 (Springer, Berlin Heidelberg 2008).
\bibitem{PN} P. Grelu and N. Akhmediev, Nat. Photonics 6, 84 (2012).
\bibitem{SR5} C. Liu, Y. Ren, Z. Y. Yang, and W. L. Yang, \href{https://aip.scitation.org/doi/abs/10.1063/1.4999916}{Chaos \textbf{27}, 083120 (2017)}.
\bibitem{SR6} Y. Ren, X. Wang, C. Liu, Z. Y. Yang, and W. L. Yang, \href{https://www.sciencedirect.com/science/article/pii/S1007570418300947}{Commun. Nonlinear. Sci. Numer. Simulat. \textbf{63}, 161 (2018)}.
\bibitem{CLIP} C. Liu, Z. Y. Yang, W. L. Yang, and N. Akhmediev, \href{https://www.osapublishing.org/josab/abstract.cfm?uri=josab-36-5-1294}{J. Opt. Soc. Am. B 36, 1294 (2019)}.
\end{thebibliography}
\end{document}